\pdfoutput=1
\documentclass[11pt]{article}

\usepackage[preprint]{acl}

\usepackage{times}
\usepackage{latexsym}

\usepackage[T1]{fontenc}
\usepackage[utf8]{inputenc}

\usepackage{microtype}

\usepackage{inconsolata}

\usepackage{graphicx}
\usepackage{longtable}
\usepackage{colortbl}
\usepackage{multirow}
\usepackage{arydshln}
\usepackage{amsmath}
\usepackage{xcolor}
\usepackage{array}
\usepackage{booktabs}
\usepackage{makecell}
\usepackage{tabularx}
\usepackage{algorithm}
\usepackage{algorithmic}
\usepackage{amssymb}

\title{From Prediction to Intervention:\\Personalized Meal-Level Glucose Regulation via an LLM Agent}

\author{Mingyu Huang$^{1,2}$, Weiqing Min$^{1,2}$, Ying Jin$^{1,2,3}$, Yilin Wang$^{1,2}$, Shuqiang Jiang$^{1,2}$\footnotemark[1] \\
   $^{1}$State Key Laboratory of AI Safety, Institute of Computing Technology,\\ Chinese Academy of Sciences, Beijing, China. \\$^{2}$University of Chinese Academy of Sciences, Beijing, China.\\
   $^{3}$School of Advanced Interdisciplinary Sciences, University of Chinese\\ Academy of Sciences, Beijing, China.\\
   \texttt{huangmingyu181@mails.ucas.ac.cn, sqjiang@ict.ac.cn} \\
    }

\begin{document}
\maketitle
\renewcommand{\thefootnote}{\fnsymbol{footnote}}
\footnotetext[1]{Corresponding Author.}

\begin{abstract}
Personalized glucose regulation remains a central yet unresolved challenge in precision nutrition, as postprandial glucose response varies substantially across individuals. Existing approaches based on glycemic indices fail to adequately account for such heterogeneity and lack the mechanism to dynamically adjust meals based on personal physiological feedback. In this context, recent advances in LLM-based agents offer a promising direction, as they enable context-aware reasoning and iterative refinement. 
Inspired by this, we propose a physio-feedback agentic loop, a unified system that integrates individualized absorption modeling with dietary intervention to regulate glucose response.
Specifically, we develop a Physiology-Aware Glucose Predictor to model individualized absorption dynamics through a learnable Temporal Physiological Absorption Decay Module. We then construct a Prediction-Driven Two-Stage Meal Optimization Agent that iteratively refines real-world meals using predicted outcomes as explicit feedback. Through extensive experiments on multiple public datasets, we demonstrate that our method not only improves prediction accuracy but also effectively reduces glucose excursions. To the best of our knowledge, this paper marks the first step in integrating physiological learning with an LLM-based agent for personalized glucose regulation.
\end{abstract}

\section{Introduction}

\begin{figure}[htbp]
\centering
\includegraphics[width=1\linewidth]{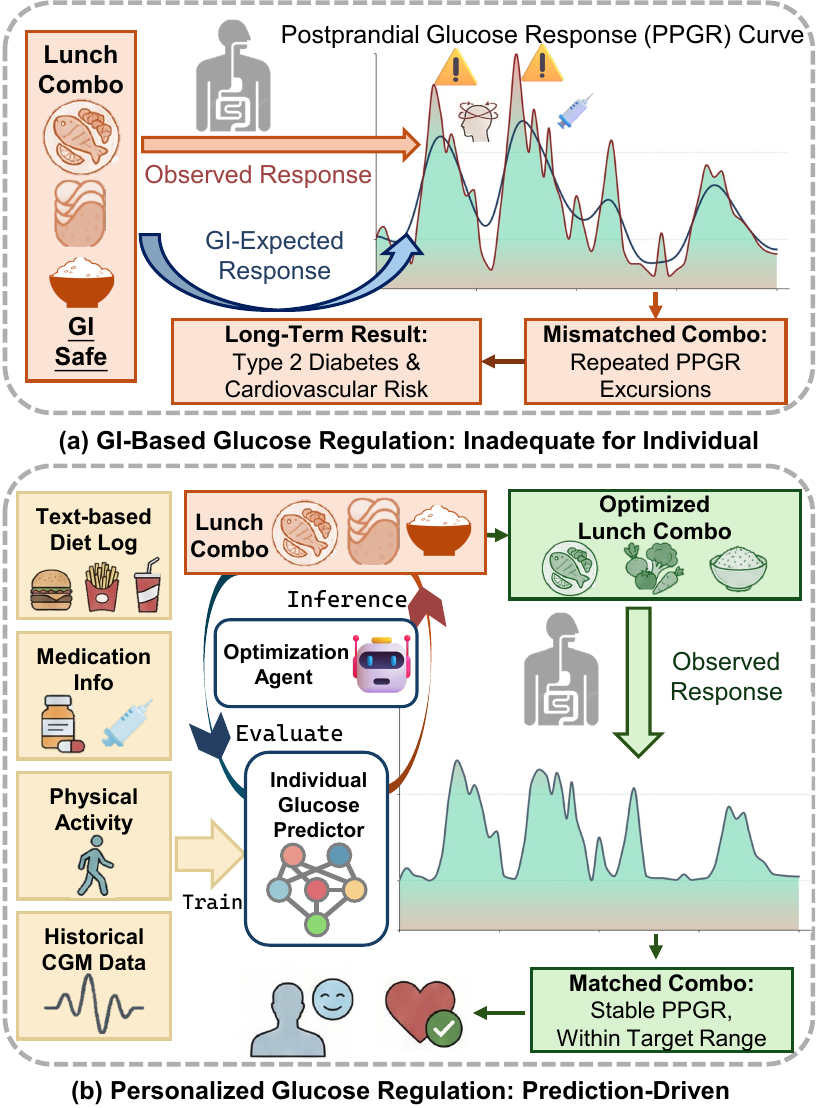}
\caption{From GI-Based Population-Level Guidance to Personalized, Physiology-Aware Glucose Regulation}
\label{fig:intro}
\end{figure}

Maintaining a stable Post-Prandial Glucose Response (PPGR) is a challenge in precision nutrition and metabolic health, as excessive glucose excursions are strongly associated with acute and chronic symptoms such as oxidative stress, vascular damage, and increased cardiovascular risk~\cite{Monnier2006OxStress, Chen2022GV}. Crucially, PPGRs to identical meals vary substantially across individuals for differences in physiology, lifestyle, and metabolic state. A large-scale study from Cell has demonstrated that population-level Glycemic Indices (GI) fail to capture this heterogeneity shown in Figure~\ref{fig:intro} (a), and that effective glucose control requires individualized strategies~\cite{zeevi2015}. Following this evidence, PPGR control requires a personalized glucose regulation system: given an individual’s physiological and contextual factors, the system should anticipate the PPGR and recommend meal adjustments tailored to it.

Despite growing interest in PPGR control, most existing approaches remain generic and fail to achieve true personalization. Dietary recommendation and meal planning methods for glucose regulation typically rely on heuristic rules and glycemic indices~\cite{Khamesian_2025, wang2025dynamic}. These approaches operate in a static manner and are largely agnostic to individual-specific glucose dynamics, resulting in limited adaptability across individuals. Meanwhile, the widespread adoption of Continuous Glucose Monitoring (CGM) systems has enabled fine-grained observation of glucose dynamics, motivating extensive research on predicting future PPGR from diet, physical, and medication records~\cite{lim2025deep, singh2025personalized, tominaga2025transformer, brugger2025ppg}. While these models have achieved encouraging forecasting accuracy, they are rarely integrated into dietary decision making. As a result, personalized PPGR regulation remains elusive due to the lack of prediction–intervention integration. Accurate glucose prediction alone does not translate into effective regulation without being directly coupled to meal-level decision making, while dietary optimization without individualized physiological feedback cannot adapt to personal glucose dynamics. Therefore, prediction and intervention should be treated as \textbf{a unified, closed-loop process} for personalized PPGR regulation.

Achieving such process requires a decision-making mechanism reasoning at meal level, adapting to individual physiological responses, and updating interventions based on predicted outcomes. Recent advances in Large Language Model (LLM) agents provide a promising foundation in this regard, as they support constraint-aware decision making and multi-step refinement~\cite{lin2024decision}. However, leveraging LLM agents for personalized glucose regulation is not straightforward because generic agent reasoning that relies on population is insufficient to capture the context-dependent meal-level decisions. The first challenge is that an agent without access to individualized glucose predictions ignores personal physiological heterogeneity, leading to mismatched recommendations for a given individual. The second challenge is that meal planning is a combinatorial optimization problem, where a one-shot recommendation is unlikely to satisfy dietary constraints while achieving the desired PPGR targets~\cite{lee2021diet,lee2021mind}. 
% An effective agent must therefore operate within a closed-loop setting.

In this work, we propose an integrated method that combines physiology-aware PPGR prediction with prediction-guided iterative meal optimization to address the above challenges, as illustrated in Figure~\ref{fig:intro} (b). At the prediction level, we introduce the \textbf{Physiology-Aware Glucose Predictor (PAGP)}, which is developed subject-specifically to capture individual glucose dynamics. PAGP explicitly models individualized food and medication absorption through a learnable \textbf{Temporal Physiological Absorption Decay Module (TPADM)} grounded in established physiological principles. Through decomposed signal modeling and TPADM, PAGP produces accurate PPGR predictions that reflect unique physiological characteristics, providing a reliable foundation for personalized intervention. At the intervention level, we propose the \textbf{Prediction-Driven Two-Stage Meal Optimization Agent (PD-2SMO)}, which performs optimization at an individual’s actual meal level. Conditioning on subject-specific glucose predictions from PAGP, the agent leverages LLM-based reasoning to iteratively adjust meal compositions using predicted PPGRs as explicit feedback. This two-stage, iterative design enables interpretable and individualized optimization of real-world meals composed of multiple dishes, allowing dietary recommendations to be continuously adapted to the predicted PPGR.

Extensive experiments across multiple public real-world datasets demonstrate that our approach not only improves PPGR prediction accuracy, achieving reductions from $20.03$ to $13.24$ in RMSE on the typical Shanghai T2DM dataset, but also effectively reduces PPGR excursions through optimized meal plans. In particular, the proposed agent leads to a decrease from $170.31$ to $142.69$ in incremental area under the PPGR curve (BIG IDEAS dataset) relative to the baseline. 

Our main contributions are threefold: 
\begin{itemize}
    \item We formulate personalized postprandial glucose regulation as a unified prediction–intervention problem, explicitly linking subject-specific PPGR prediction with meal-level dietary decision making via an LLM agent.
    \item We propose an integrated method that combines the PAGP, which models individualized absorption dynamics through a learnable physiological module, with the PD-2SMO agent that iteratively optimizes real-world meals using predicted PPGRs as feedback.
    \item We provide a comprehensive evaluation across multiple real-world datasets, demonstrating that our method improves both PPGR prediction accuracy and regulation performance compared with baselines.
\end{itemize}

\section{Related Work}
The prediction of PPGR has evolved from population-based to personalized models. Recent large-scale studies have elucidated that PPGR are driven by complex interactions between diet, microbiome, and host physiology~\cite{zeevi2015}. Individual variations in glycemic responses to specific macronutrients are deeply rooted in underlying metabolic physiology, necessitating subject-specific modeling approaches~\cite{wu2025}. With the proliferation of CGM, data-driven approaches have gained much attention. Machine learning models have been deployed to identify metabolic subphenotypes of Diabetes~\cite{Metwally2025Prediction}. However, pure data-driven models often lack physiological interpretability. To bridge this gap, the concept of Medical Digital Twins has emerged as a paradigm for integrating physiological knowledge with data-driven predictions~\cite{Marchal2025Applications}.

While prediction is a prerequisite, the ultimate goal of glucose management is regulation. Extensive research has been conducted on the Artificial Pancreas (AP), which automates insulin delivery for PPGR regulation. These studies focus on clinical models, fuzzy logic models and machine learning models for AP systems~\cite{hettiarachchi2022integrating, tejedor2020reinforcement}. However, while AP systems manage glucose through pharmacological intervention, controlling PPGR via dietary lifestyle intervention remains a distinct and equally critical imperative. Existing dietary recommendation systems lacks the closed-loop feedback mechanism found in AP systems. LLMs have demonstrated remarkable capabilities in planning and loop-reasoning, extending their utility from text generation to autonomous decision-making agents~\cite{he-etal-2025-pasallmagent,zhu-etal-2025-plangptllmagent}. In the healthcare domain, LLMs are increasingly applied to complex tasks such as diagnostic reasoning, patient triage, and lifestyle coaching~\cite{du-etal-2025-llmagentSP}. Specifically for nutrition, LLMs offer a flexible interface for processing unstructured dietary information and generating recipes. However, standard LLMs often hallucinate nutritional content or fail to strictly adhere to numerical constraints. In the context of diet, collaborative agent frameworks have been proposed to simulate user-nutritionist interactions for weight loss plans~\cite{Li2024RolePlay}. Nevertheless, these agents typically operate on general nutritional guidelines rather than personalized physiological feedback.

\section{Methods}
\begin{figure*}[htbp]
\centering
\includegraphics[width=0.8\linewidth]{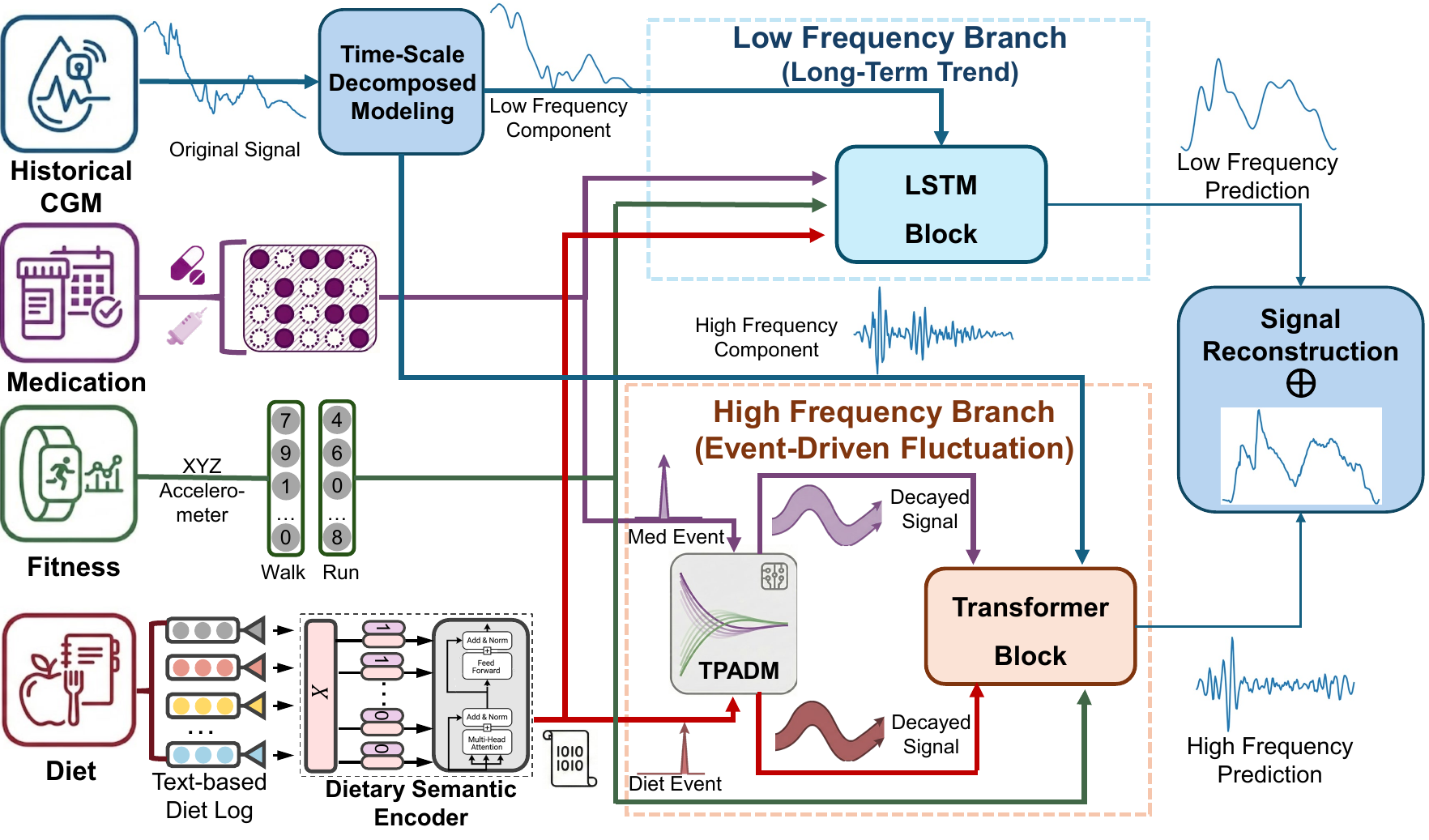}
\caption{Overview of the Physiology-Aware Glucose Predictor. Historical CGM is decomposed into low-/high-frequency components via variational mode decomposition; an LSTM models long-term trends, while a Transformer and Temporal Physiological Absorption Decay Module model event-driven fluctuations; predicted components are reconstructed to form the final PPGR.}
\label{fig:model}
\end{figure*}
\label{sec:method}
Personalized PPGR regulation requires capturing individual glucose dynamics and translating such predictions into feasible interventions. To achieve this, we propose a unified method that tightly couples personalized glucose prediction with prediction-guided meal optimization. Our method consists of two parts. In Section~\ref{sec:method:pagp}, we introduce how to forecast individualized PPGRs by incorporating physiological representations of dietary and medication absorption. In Section~\ref{sec:method:tsimoa}, we illustrate how to generate an optimized meal configuration that improves postprandial glycemic outcomes given the corresponding user context.

% ============================================================
\subsection{Personalized Glucose Prediction}
\label{sec:method:pagp}

We propose the \textbf{Physiology-Aware Glucose Predictor (PAGP)} to predict future glucose trajectories given personal historical CGM data, dietary records, physical activity, and medication information. 
The key innovations of PAGP are threefold:
(i) explicit decomposition of glucose dynamics into heterogeneous time scales,
(ii) individual event-driven high-frequency modeling under physiological absorption constraints, and
(iii) end-to-end integration of dietary semantics with structured physiological signals. Figure~\ref{fig:model} provides an overview of the proposed PAGP.

% ------------------------------------------------------------
\subsubsection{Time-Scale Decomposed Modeling}

Glucose dynamics are governed by physiological mechanisms operating at different temporal scales. 
Slowly variations reflect basal metabolism, insulin sensitivity, and circadian rhythms, while rapid fluctuations are driven by discrete events such as meals and medication intake. Modeling these heterogeneous dynamics using a single predictor often leads to interference between long-term trend learning and short-term event modeling.

To address this issue, we apply Variational Mode Decomposition (VMD)~\cite{Dragomiretskiy2014VMD, Wang2020VMDLSTM} to decompose the historical glucose signal $G(t)$ into band-limited intrinsic mode functions, which are further aggregated into low- and high-frequency components:
\begin{equation}
G(t) = G_L(t) + G_H(t)
= \sum_{k \in \mathcal{L}} u_k(t) + \sum_{k \in \mathcal{H}} u_k(t),
\end{equation}
where $\mathcal{L}$ and $\mathcal{H}$ denote the sets of low- and high-frequency modes, respectively. During inference, the predicted low- and high-frequency components are reconstructed in the time domain to obtain the final glucose prediction.

The low-frequency component $G_L(t)$ captures smooth, slowly evolving glucose trends and is modeled using an LSTM-based predictor:
\begin{multline}
\hat{G}^{\text{low}}(t+1:t+H) = \\
f_{\text{LSTM}}\!\left(
G^{\text{low}}(1:t),
\mathbf{e}^{\text{meal}}_{1:t},
\mathbf{x}^{\text{act}}_{1:t},
\mathbf{x}^{\text{med}}_{1:t}
\right),
\end{multline}
where $\mathbf{x}^{\text{act}}$ and $\mathbf{x}^{\text{med}}$ denote structured activity and medication features, respectively.
This branch is supervised exclusively by the VMD-derived low-frequency signal, enabling the model to focus on long-term glucose evolution without being affected by event-driven fluctuations.

% ------------------------------------------------------------
\subsubsection{Event-Driven Fluctuation}

Rapid glucose fluctuations are predominantly induced by absorption event and medication event.
To explicitly model these physiological mechanisms, we introduce the \emph{Temporal Physiological Absorption Decay Module (TPADM)} for high-frequency glucose prediction.

TPADM is constructed based on analytical solutions of differential equations derived from the Hovorka physiological model~\cite{Hovorka2004MPC}.
A first-order absorption or action process is expressed as:
\begin{equation}
\frac{d s(t)}{dt} = -\frac{1}{\tau} s(t) + \kappa \, u(t),
\end{equation}
with the analytical solution:
\begin{equation}
s(t) = s(0)\, e^{-t/\tau} + \kappa \int_{0}^{t} e^{-(t-\xi)/\tau} u(\xi)\, d\xi,
\end{equation}
where $u(t)$ represents dietary or medication input, $s(t)$ denotes the physiological effect state, and $\tau$ and $\kappa$ control temporal decay and effect magnitude. $s(0)$ represents the residual absorption state from meals consumed prior to the current context window. In our implementation, we initialize $s(0)$ assuming the window starts after the absorption of distant previous meals has subsided. In TPADM, both $\tau$ and $\kappa$ are treated as individual learnable parameters and optimized end-to-end, allowing the model to adapt to individual-specific absorption and action dynamics.

While the analytical solution operates on a continuous input $u(t)$, real-world dietary and medication logs are discrete. In our formulation, a discrete meal event at time $t_i$ with carbohydrate content $D_i$ is modeled mathematically as an impulse in the continuous domain: $u(t) = \sum D_i \cdot \delta(t-t_i)$. When this impulse input is applied to the above equation, the integral yields the analytical decay curve. The medication intakes are treated as the same. These continuous physiological states are then discretely sampled at step $t$ to form the physiological driving sequence:
\begin{equation}
\mathbf{z}_{1:t} =
f_{\text{TPADM}}\!\left(
\mathbf{e}^{\text{meal}}_{1:t},
\mathbf{x}^{\text{med}}_{1:t}
\right).
\end{equation}

The physiological driving sequence is then fed into a Transformer to predict the high-frequency glucose component:
\begin{multline}
\hat{G}^{\text{high}}(t+1:t+H)
=\\
f_{\text{Trans}}\!\left(
G^{\text{high}}(1:t),
\mathbf{x}^{\text{act}}_{1:t},
\mathbf{z}_{1:t}
\right).
\end{multline}
This branch is supervised solely by the VMD-derived high-frequency signal, enabling focused modeling of rapid, event-driven glucose responses under explicit physiological constraints.

% ------------------------------------------------------------
\subsubsection{Dietary Semantic Encoder}

PAGP integrates both structured and unstructured inputs.
Historical CGM, physical activity, and medication records are treated as structured time-series signals, while dietary descriptions are provided in natural language form.
Dietary information is encoded using a nutrition-aware semantic encoder based on a BERT architecture~\cite{devlin-etal-2019-bert}
which is trained jointly with the overall model to capture task-relevant dietary semantics associated with PPGRs. Both low- and high-frequency branches are optimized end-to-end under branch-specific supervision.

% ============================================================
\subsection{Meal Optimization}
\label{sec:method:tsimoa}

\begin{figure*}[htbp]
\centering
\includegraphics[width=1\linewidth]{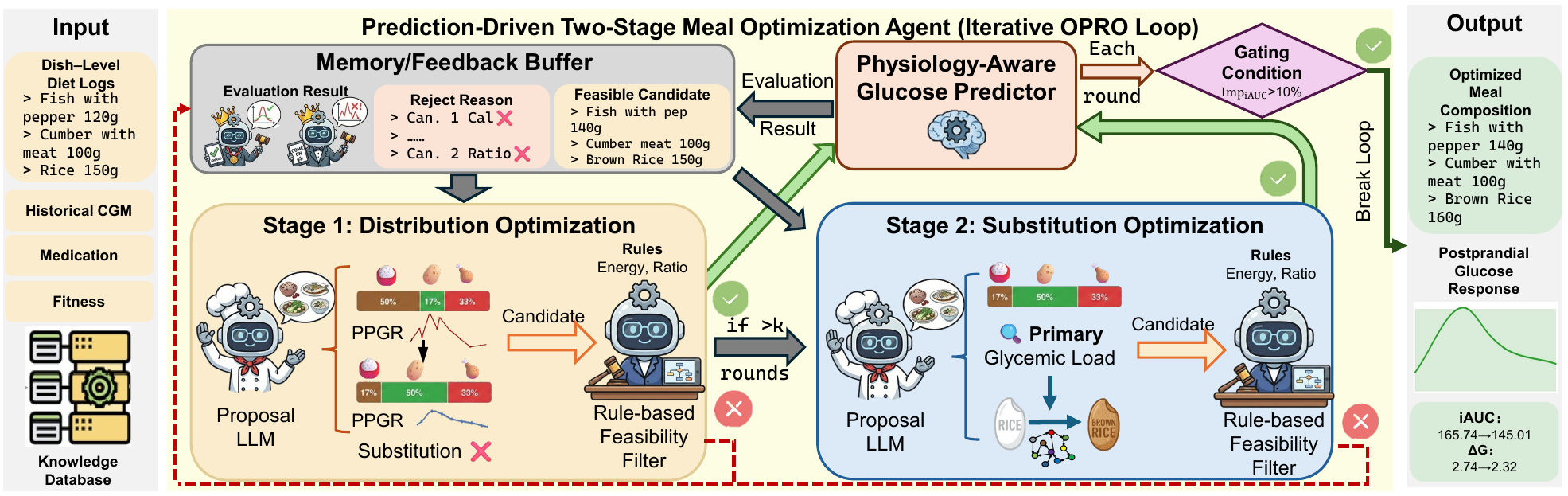}
\caption{The Prediction-Driven Two-Stage Meal Optimization Agent iteratively proposes meal candidates via optimization by prompting, enforces rule-based feasibility constraints, and uses a personalized predictor to score candidates. It first performs distribution-only adjustment and enables minimal ingredient substitution only when needed; optimization targets include incremental area under the curve iAUC$_{2h}$ and peak glucose increment $\Delta G_{2h}$.}
\label{fig:agent}
\end{figure*}

Building upon PAGP, we propose the \textbf{Prediction-Driven Two-Stage Meal Optimization Agent (PD-2SMO)} to generate an optimized meal configuration that improves PPGR outcomes while remaining feasible in real-world dietary practice.
The core innovations of PD-2SMO include:
(i) a distribution-first, substitution-on-demand optimization strategy, and
(ii) a iterative agent treating PPGR prediction as an optimization guider. Figure~\ref{fig:agent} provides an overview of the proposed PD-2SMO.

The PAGP described in Section~\ref{sec:method:pagp}
is directly reused as a personalized \emph{prediction-based evaluator} for meal optimization.
Given user-specific contextual information $u$ and a candidate meal $x$, PAGP deterministically
predicts the corresponding 2-hour PPGR. For a certain participants, all candidate meals are assessed using the same personalized predictor, ensuring consistent and
individualized evaluation across optimization iterations.

% ------------------------------------------------------------
\subsubsection{Two-Stage Strategy}

Unlike conventional dietary recommendation systems that directly substitute food items, PD-2SMO adopts a \emph{distribution-first} optimization strategy grounded in real-world dietary behavior.

In \textbf{Stage~1}, the agent is restricted to adjusting the distribution and proportions of existing meal components while keeping ingredient types fixed.
Hard constraints are enforced, including:
(i) total energy deviation within $\pm10\%$ of the original meal,
(ii) no seasoning- or spice-level modification, and
(iii) no substitution of core ingredients.
If the target improvement is achieved at this stage, the optimization terminates.

Only when distribution-level adjustment fails does the agent proceed to \textbf{Stage~2}, where minimal ingredient substitution is allowed.
Substitutions are restricted to a small candidate pool (e.g., staple-to-staple replacement) to ensure minimal deviation from the original meal.
This \emph{distribution-first, substitution-on-demand} gating mechanism constitutes a central novelty of our approach.

% ------------------------------------------------------------
\subsubsection{OPRO-Based Iterative Optimization}

PD-2SMO is implemented using Optimization by PROmpting (OPRO)~\cite{yang2024large} with the proposed PAGP providing feedback signals.
At iteration $t$, the agent constructs a prompt containing the original meal, user context, stage-specific constraints, and historical feedback.
The LLM generates a set of candidate meals:
\begin{equation}
C_t = \{x^{(t)}_1, \dots, x^{(t)}_K\} \sim \Pi(\cdot \mid P_t).
\end{equation}
Candidate meals are first passed through a rule-based feasibility filter that enforces energy, executability, and safety constraints.
Infeasible candidates are discarded, while feasible ones are evaluated by the PAGP.
The agent maintains a memory buffer of top-performing candidates and recent failures, which guides subsequent prompt construction.
This iterative loop continues until the optimization target is met or the evaluation budget is exhausted.

\section{Experimental Evaluation}

\subsection{Dataset Overview}

This study encompasses four representative datasets for glucose prediction and personalized nutrition research: Shanghai T1DM, Shanghai T2DM~\cite{zhao2023chinese}, CGMacros~\cite{hossain2025cgmacros}, and the BIG IDEAS~\cite{bent2021bigidea} dataset. 
Together, these datasets cover a broad demographic and physiological spectrum, ranging from healthy individuals and those with prediabetes to patients with Type~1 and Type~2 diabetes. All datasets provide multimodal data, including CGM time-series and contextual information such as dietary logs, insulin administration records, medication usage, and physical activity records.
An overview of dataset composition is summarized in Table~\ref{tab:datasets_summary}.

\begin{table}[htbp]
  \centering
  \small
  \begin{tabularx}{\columnwidth}{@{}l c c c X@{}}
    \toprule
    \textbf{Dataset} & \textbf{H/Pre} & \textbf{T1} & \textbf{T2} & \textbf{Composition} \\
    \midrule
    SH T1 & 0 & 12 & 0 & CGM(15'), Ins., Diet \\
    SH T2 & 0 & 0 & 100 & CGM(15'), Diet, Med. \\
    CG & 31 & 0 & 14 & CGM(1'), Diet, Fit, Med. \\
    IDEAS & 16 & 0 & 0 & CGM(5'), Diet, Fit \\
    \bottomrule
  \end{tabularx}
\caption{Dataset overview. We report participant composition across healthy / pre-diabetes / type-1 / type-2 diabetes (H/Pre/T1/T2) and available modalities. CGM sampling intervals are shown in minutes (e.g., 15'). Abbreviations: SH T1: Shanghai T1DM; SH T2: Shanghai T2DM; CG: CGMacros; IDEAS: BIG IDEAS; Ins.: insulin injection; Fit.: physical activity; Med.: medication.}
\label{tab:datasets_summary}
\end{table}

% ============================================================
\subsection{Evaluation of PAGP}

For each participant, an individual PAGP model is trained and evaluated, following a three-stage protocol:
\textbf{(1) Population-level pretraining.}  
The model is pretrained on participants with the same physiological condition as the target subject. 
\textbf{(2) Individual fine-tuning.}  
The pretrained model is fine-tuned using the training split of the target individual. 
\textbf{(3) Individual evaluation.}  Performance is evaluated on the held-out test set of the same individual.
This procedure yields one evaluation result per participant. Final performance is reported as the average across all individuals in each dataset, ensuring that results reflect personalized prediction capability rather than population-level fitting.

The prediction task focuses on postprandial glucose forecasting over a \textbf{120-minute horizon} following meal intake.
Given historical CGM data and contextual inputs up to time $t$, models predict the future glucose trajectory:
\[
\hat{G}(t+1:t+120).
\]
Prediction performance is primarily evaluated using the \textbf{Root Mean Square Error ($RMSE$)} which is computed per individual and then averaged across participants. In addition, we assess the agreement between predicted and observed glucose trajectories using the coefficient of determination ($R^2$). The calculation of $RMSE$ and $R^2$ is detailed in Appendix~\ref{PAGPMetric}. While RMSE captures absolute prediction error, $R^2$ reflects how well models reproduce temporal glucose variation patterns.

We compare PAGP with several representative glucose prediction baselines, including LSTM~\cite{hochreiter1997long}, GluNet~\cite{li2020glunet}, GlucoNet~\cite{farahmand2024gluconet}, and GluFormer~\cite{sergazinov2023gluformer}.
All baseline models follow the same personalized training protocol to ensure a fair comparison.

\begin{figure}[htbp]
\centering
\includegraphics[width=0.8\linewidth]{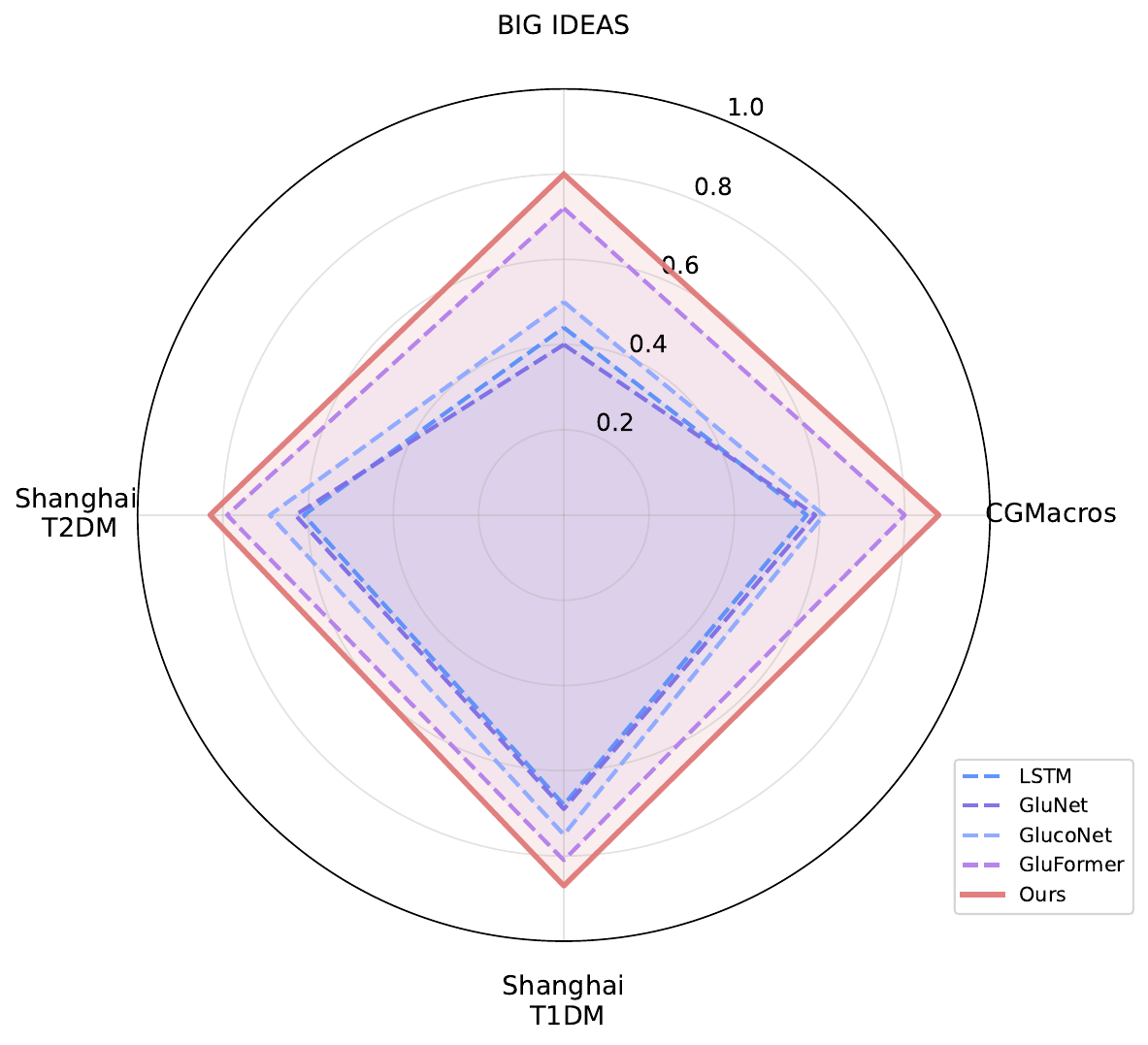}
\caption{Radar chart of cross-dataset prediction agreement measured by the coefficient of determination $R^2$. Higher $R^2$ indicates better alignment between predicted and observed PPGRs across datasets.}
\label{fig:radar_r2}
\end{figure}

% ------------------------------------------------------------
\subsubsection{Results and Cross-Dataset Comparison}
\begin{table}[htbp]\small
\centering
\begin{tabular}{l c c c c}
\toprule
Methods & SH T1 & SH T2  & CG & IDEAS   \\
\midrule
LSTM     & 28.22 & 21.56    & 23.37 & 17.79  \\
GluNet   & 26.90 & 20.03    & 23.01 & 19.28  \\
GlucoNet & 25.19 & 17.16   & 21.67 & 16.73   \\
GluFormer & 20.73 & 14.54   & 16.91 & 12.94  \\
\midrule
Ours     & \textbf{16.97} & \textbf{13.24}  & \textbf{11.39} & \textbf{10.39}   \\
\bottomrule
\end{tabular}
\caption{PPGR forecasting performance over a 120-minute horizon reported in RMSE across datasets. Lower RMSE is better. Best results are in bold.}
\label{tab:rmse_comparison}
\end{table}

Table~\ref{tab:rmse_comparison} summarizes the average RMSE of different models for postprandial glucose prediction over a 120-minute horizon across five benchmark datasets. Overall, PAGP achieves the lowest RMSE on all datasets, demonstrating consistently improved prediction accuracy compared with both recurrent and Transformer-based baselines.

Specifically, PAGP reduces RMSE from 12.94 to 10.39 on the IDEAS dataset when compared with the strongest baseline GluFormer. On the CGMacros dataset, PAGP achieves an RMSE of 11.39, representing a substantial improvement over GluFormer (16.91) and earlier convolutional or recurrent models. Similar trends are observed on the Shanghai T1 and T2DM datasets, where PAGP attains RMSEs of 16.97 and 13.24, respectively, outperforming GluFormer by notable margins. These results suggest that PAGP generalizes well across heterogeneous populations and data collection settings, including both controlled research cohorts and real-world clinical datasets. We further conduct paired $t$-tests on prediction results against the strongest baseline, GluFormer. 
Across all four datasets, the RMSE reduction achieved by PAGP is statistically significant ($p<0.01$), and the improvement in $R^2$ is also statistically significant ($p<0.05$). 

\begin{table}[htbp]\small
\centering
\begin{tabular}{c c |  c c c c}
\toprule
Emb & TPADM  & IDEAS & CG & SH T1 & SH T2 \\
\midrule
$\boldsymbol{\times}$ & $\checkmark$  & 13.35 & 16.20 & 19.04 & 16.67 \\
$\checkmark$ & $\boldsymbol{\times}$  & 11.92 & 17.23 & 17.91 & 16.10 \\
$\checkmark$ & $\checkmark$  & 10.39  & 11.39 & 16.97 & 13.24 \\
\bottomrule
\end{tabular}
\caption{Ablation of PAGP components on PPGR forecasting reported in RMSE. Emb denotes the dietary semantic encoder. Best results are in bold.}
\label{tab:rmse_ablation}
\end{table}

To further assess robustness across heterogeneous datasets, Fig.~\ref{fig:radar_r2} presents a radar chart of averaged $R^2$ scores.
Each axis corresponds to one dataset, highlighting the relative stability and generalization ability of different models.

% ------------------------------------------------------------
\subsubsection{Ablation Study}

We conduct ablation experiments to analyze the contribution of key model components. Specifically, we remove the BERT-based dietary semantic encoder and the TPADM module, respectively. In the semantic encoder removed scenario, we quantize the text-based dietary logs to structured calorie, carbohydrate, fat and protein logs as our model input. As shown in Table~\ref{tab:rmse_ablation}, removing either component leads to noticeable performance degradation, confirming the importance of both dietary semantic encoding and physiology-inspired absorption modeling.

% ============================================================
\subsection{Evaluation of PD-2SMO}
\begin{table*}[htbp]
\small
\centering
\setlength{\tabcolsep}{3.5pt}
\renewcommand{\arraystretch}{1.1}
\begin{tabular}{lcccccc|cccccc}
\toprule
\multirow{3}{*}{Dataset}
& \multicolumn{6}{c}{Main Results}
& \multicolumn{6}{c}{GluFormer-based Generalization} \\
\cmidrule(lr){2-7}
\cmidrule(lr){8-13}
& \multicolumn{2}{c}{Original}
& \multicolumn{2}{c}{ReAct}
& \multicolumn{2}{c}{Ours}
& \multicolumn{2}{c}{Original}
& \multicolumn{2}{c}{ReAct}
& \multicolumn{2}{c}{Ours} \\
\cmidrule(lr){2-3}\cmidrule(lr){4-5}\cmidrule(lr){6-7}
\cmidrule(lr){8-9}\cmidrule(lr){10-11}\cmidrule(lr){12-13}
& iAUC & $\Delta G$ & iAUC & $\Delta G$ & iAUC & $\Delta G$
& iAUC & $\Delta G$ & iAUC & $\Delta G$ & iAUC & $\Delta G$ \\
\midrule
BIG IDEAS & 214.78 & 3.03 & 170.31 & 2.57 & \textbf{142.69} & \textbf{2.04}
          & 229.04 & 3.08 & 193.67 & 2.76 & \textbf{159.83} & \textbf{2.49} \\
CGMacros  & 225.57 & 2.98 & 189.65 & 2.64 & \textbf{158.72} & \textbf{2.39}
          & 243.75 & 3.01 & 201.73 & 2.80 & \textbf{176.37} & \textbf{2.61} \\
Shanghai T1DM & 140.35 & 2.13 & 132.86 & 2.01 & \textbf{121.95} & \textbf{1.76}
               & 154.81 & 2.34 & 140.39 & 2.12 & \textbf{134.35} & \textbf{2.01} \\
Shanghai T2DM & 163.67 & 2.55 & 151.84 & 2.30 & \textbf{137.09} & \textbf{2.06}
               & 169.52 & 2.58 & 152.90 & 2.34 & \textbf{141.58} & \textbf{2.12} \\
\bottomrule
\end{tabular}
\caption{Meal optimization results evaluated by iAUC$_{2h}$ and $\Delta G_{2h}$ (lower is better). Results using the proposed PAGP for evaluation. Right: generalization when replacing the evaluator with GluFormer. Baselines include the original meal and a ReAct agent. Best results are in bold.}
\label{tab:main_and_generalization}
\end{table*}

We evaluate PD-2SMO on composed meals from multiple datasets.
Each lunch or dinner is treated as an independent optimization instance.
For each meal, the agent generates an optimized meal plan and predicts the corresponding postprandial glucose trajectory over a 120-minute window. Evaluation mainly focuses on two clinical metrics:
\textbf{(i) incremental area under the glucose curve, iAUC$_{2h}$, and
(ii) maximum glucose increment, $\Delta G_{2h} = \max_{t \in [0,2h]}(G(t)-G(0))$.}
Additional results, including time in range, time in hyperglycemia and time in hypoglycemia, are reported in Appendix~\ref{tir_result}. 

We compare PD-2SMO against two baselines:(i) the original meal composition, and (ii) a general Reason+Act (ReAct) style agent~\cite{yao2023react}. To ensure a fair comparison, both ReAct and PD-2SMO are run in the same optimization environment, using the same personalized PAGP evaluator and the same rule-based feasibility constraints.

% ------------------------------------------------------------
\subsubsection{Optimization Results and Comparison}
Table~\ref{tab:main_and_generalization} reports the meal optimization results across four datasets. Overall, PD-2SMO consistently outperforms both the original meals and baseline strategies, achieving substantial reductions in postprandial glucose exposure and peak glucose excursions.
Compared with original meals, PD-2SMO achieves pronounced iAUC reductions across all datasets, decreasing iAUC from $214.78$ to $142.69$ on BIG IDEAS, and from $163.67$ to $137.09$ on Shanghai T2DM. Across all datasets, reductions in $\Delta G_{2h}$ align with iAUC reductions, suggesting that PD-2SMO not only lowers overall PPGR exposure but also mitigates glucose spikes. These results demonstrate that PD-2SMO delivers effective meal-level glucose optimization across heterogeneous populations and dietary contexts. Moreover, PD-2SMO further yields consistent additional gains compared with the ReAct agent, producing lower glucose excursions across both evaluation metrics. This shows the effectiveness of our proposed mechanism. We further conduct paired $t$-tests on optimization results against the ReAct baseline. Across all datasets involved, the iAUC reductions from ReAct to our method are statistically significant ($p < 0.01$). The $\Delta G_{2h}$ reductions from ReAct to ours are also statistically significant ($p<0.01$).

% ------------------------------------------------------------
\subsubsection{Generalization Across Predictors and LLM Backbones}
The PD-2SMO is developed on the Qwen3-Max LLM backbone~\cite{yang2025qwen3technicalreport}. To examine whether it depends on a specific glucose predictor or LLM, we further evaluate performances under alternative predictors and LLM backbones. As shown in Table~\ref{tab:main_and_generalization} (right) and Table~\ref{tab:generalization_llms}, PD-2SMO consistently preserves its performance across both dimensions. Replacing the PAGP with GluFormer results in reductions in both iAUC and $\Delta G$ across all datasets, closely matching the performance observed with the PAGP. Although absolute metric values vary for differences in predictor behavior, the PD-2SMO consistently outperform both original meals and ReAct-based optimization. This suggests that PD-2SMO is predictor-agnostic rather than relying on a specific forecasting model. Table~\ref{tab:generalization_llms} evaluates PD-2SMO under three different LLM backbones. The iAUC and $\Delta G$ values remain stable and comparable, with no single backbone dominating performance.
Together, these results demonstrate that PD-2SMO generalizes well across heterogeneous glucose predictors and LLM backbones, supporting its applicability as a modular and model-agnostic meal optimization framework.

\begin{table}[htbp]
\small
\centering
\setlength{\tabcolsep}{4pt}
\renewcommand{\arraystretch}{1.1}
\begin{tabular}{lcccccc}
\toprule
Dataset
& \multicolumn{2}{c}{Deepseek-v3.2}
& \multicolumn{2}{c}{Gemini 3 Pro}
& \multicolumn{2}{c}{GPT-4o} \\
\cmidrule(lr){2-3}\cmidrule(lr){4-5}\cmidrule(lr){6-7}
& iAUC & $\Delta G$
& iAUC & $\Delta G$
& iAUC & $\Delta G$ \\
\midrule
IDEAS     & 151.56 & 2.10 & 140.81 & 2.02 & 146.97 & 2.09 \\
CG      & 163.19 & 2.45 & 151.89 & 2.31 & 159.81 & 2.39 \\
SH T1 & 131.02 & 1.81 & 124.57 & 1.77 & 130.67 & 1.80 \\
SH T2 & 140.38 & 2.09 & 141.02 & 2.08 & 140.46 & 2.06 \\
\bottomrule
\end{tabular}
\caption{PD-2SMO Optimization performance across LLM backbones, evaluated by iAUC$_{2h}$ and $\Delta G_{2h}$ (lower is better). Results remain stable across DeepSeek-v3.2, Gemini 3 Pro, and GPT-4o.}
\label{tab:generalization_llms}
\end{table}

% ------------------------------------------------------------
\subsubsection{Ablation Study}
\begin{table*}[htbp]
\small
\centering
\setlength{\tabcolsep}{3.5pt}
\begin{tabular}{cc|cccccccc}
\toprule
\multirow{2}{*}{Subs}
& \multirow{2}{*}{Loop}
& \multicolumn{2}{c}{BIG IDEAS}
& \multicolumn{2}{c}{CGMacros}
& \multicolumn{2}{c}{Shanghai T1DM}
& \multicolumn{2}{c}{Shanghai T2DM} \\
\cmidrule(lr){3-4}\cmidrule(lr){5-6}\cmidrule(lr){7-8}\cmidrule(lr){9-10}
& & iAUC & $\Delta G$ & iAUC & $\Delta G$
& iAUC & $\Delta G$ & iAUC & $\Delta G$ \\
\midrule
\checkmark & $\boldsymbol{\times}$ & 189.83 & 2.60 & 196.37 & 2.71 & 174.35 & 2.21 & 161.58 & 2.32 \\
$\boldsymbol{\times}$ & \checkmark & 161.34 & 2.37 & 171.91 & 2.50 & 127.83 & 1.89 & 142.71 & 2.11 \\
\checkmark & \checkmark & \textbf{142.69} & \textbf{2.04}
           & \textbf{158.72} & \textbf{2.39}
           & \textbf{121.95} & \textbf{1.76}
           & \textbf{137.09} & \textbf{2.06} \\
\bottomrule
\end{tabular}
\caption{Ablation of PD-2SMO design choices reported in iAUC \& $\Delta G$ (lower is better). Subs enables the minimal ingredient substitution stage; Loop enables the iterative OPRO optimization loop. Best results are in bold.}
\label{tab:ablation_two_columns}
\end{table*}
We further study key designs through ablation and constraint-based studies. The ablation study results are summarized in Table~\ref{tab:ablation_two_columns}. The full model, which incorporates both ingredient substitution and iterative OPRO loop, consistently achieves the best performance across all datasets and evaluation metrics. Removing OPRO loop while retaining ingredient substitution leads to consistent performance degradation. Compared with the full model, iAUC rises from $121.95$ to $174,35$ on Shanghai T1DM and from $137.09$ to $161.58$ on Shanghai T2DM, with $\Delta G$ increasing accordingly. When the ingredient substitution module is removed, performance also degrades, indicating that distribution-level adjustments alone are insufficient to achieve optimal glucose regulation. The absence of either component results in higher iAUC and larger glucose excursions, demonstrating the effectiveness of ingredient substitution and OPRO loop.

To further investigate the contribution of agent-level dietary constraints, we evaluate the rationality of optimized recipe combinations after removing the calorie constraint and the joint calorie–ratio constraints, respectively. Results are reported in~\ref{const_anal}.

\section{Conclusion}
Our study presents a novel framework that bridges the gap between physiological modeling and generative artificial intelligence for precision nutrition. A critical contribution of this work is the formulation of glucose regulation as a semantic generation and combinatorial optimization problem. Our PD-2SMO leverages the semantic reasoning capabilities of LLMs to navigate this discrete action space. By adopting a distribution-first, substitution-second strategy, our method generates meal plans that are not only theoretically optimal for glucose control but also practically feasible and coherent. Furthermore, unlike static approaches based on GIs that fail to account for individual variability, our closed-loop agentic framework dynamically adapts to personal physiological responses, offering a precision medicine alternative to traditional heuristic-based diet planning.

The promising results from our offline evaluation pave the way for several future research avenues. First, while our current work validates the framework using historical data and predictive evaluation, ultimate validation requires prospective human subject trials to assess behavioral adherence and long-term metabolic outcomes in a free-living context. Second, we plan to enhance the agent's capability to handle longitudinal constraints, such as weekly nutritional balance and cost-effectiveness, moving beyond single-meal optimization to comprehensive lifestyle management. Finally, investigating rapid adaptation techniques to address the cold-start problem for new users with sparse data remains a key priority for deploying this system at scale.

\section*{Limitations}
Despite the promising results, this work has several limitations. First, the proposed framework relies on historical CGM data and individualized training or adaptation to capture personal glucose dynamics. As a result, its performance may be limited in cold-start scenarios where only sparse or short-term glucose records are available. In the future, rapid personalization strategies may help mitigate this dependency. Second, the effectiveness of the optimized meal plans is evaluated through predicted glucose responses rather than prospective human intervention studies. Real-world validation through controlled dietary experiments or longitudinal clinical trials remains necessary to assess behavioral adherence, safety, and long-term health impact. Third, although our results support the effectiveness of LLM-based meal optimization, the present study does not yet exhaustively compare against all alternative optimization paradigms, such as Bayesian optimization or other non-LLM search strategies. A broader comparison would help further isolate the unique advantages and limitations of LLM-based agents for PPGR intervention.

\section*{Ethical Considerations}
Our study is conducted with careful attention to participant privacy, informed consent, and responsible model use. The CGM-based datasets used in this work were obtained from publicly released research resources whose publishers have already applied privacy-preserving procedures at the time of release (e.g., removal of direct identifiers and appropriate de-identification). According to the dataset publishing website, all participants signed informed consent forms and received appropriate compensation for their participation, and the original data collection protocols were reviewed and approved by the corresponding institutional ethics review boards. Our work only uses the released, de-identified data and does not attempt to re-identify individuals or link records to external data sources.

Our framework includes an LLM-based meal optimization agent, which introduces additional ethical risks. First, LLMs may generate plausible but incorrect recommendations even with our constaints (i.e., hallucinations), potentially leading to unsafe dietary suggestions if used without clinical oversight. Second, the agent may inherit biases from its pretraining data and may perform unevenly across populations, diets, or cultural contexts. Third, if deployed improperly, interaction logs or user-provided context could create privacy risks. To mitigate these risks in our research setting, we position the agent as a decision-support component rather than a medical device, and we tried to constrain its outputs using predefined nutritional and feasibility rules, combined with explicit grounding in individualized glucose predictions. We also recommend that any real-world deployment should include human-in-the-loop review, additional safety filtering and continuous monitoring for errors and bias.

\section*{Acknowledgments}
This study was supported by Noncommunicable Chronic Diseases-National Science and Technology Major Project (2026ZB0556800), the Beijing Natural Science Foundation (JQ24021) and the National Natural Science Foundation of China (62125207 and 62472411).

\bibliography{glucose,ref_FoodLMM}

@inproceedings{devlin-etal-2019-bert,
    title = "{BERT}: Pre-training of Deep Bidirectional Transformers for Language Understanding",
    author = "Devlin, Jacob  and
      Chang, Ming-Wei  and
      Lee, Kenton  and
      Toutanova, Kristina",
    editor = "Burstein, Jill  and
      Doran, Christy  and
      Solorio, Thamar",
    booktitle = "Proceedings of the 2019 Conference of the North {A}merican Chapter of the Association for Computational Linguistics: Human Language Technologies, Volume 1 (Long and Short Papers)",
    month = jun,
    year = "2019",
    address = "Minneapolis, Minnesota",
    publisher = "Association for Computational Linguistics",
    url = "https://aclanthology.org/N19-1423/",
    doi = "10.18653/v1/N19-1423",
    pages = "4171--4186"
}

@article{zeevi2015,
  title={Personalized nutrition by prediction of glycemic responses},
  author={Zeevi, David and Korem, Tal and Zmora, Niv and Israeli, David and Rothschild, Daphna and Weinberger, Adina and Ben-Yacov, Orly and Lador, Dar and Avnit-Sagi, Tali and Lotan-Pompan, Maya and others},
  journal={Cell},
  volume={163},
  number={5},
  pages={1079--1094},
  year={2015},
  publisher={Elsevier},
  url= {https://www.cell.com/cell/fulltext/S0092-8674(15)0148}
}

@article{lim2025deep,
  author={Lim, Min Hyuk and Chae, Hyocheol and Yoon, Jeongwon and Shin, Insik},
  title = {A deep learning framework for virtual continuous glucose monitoring and glucose prediction based on life-log data},
  journal = {Scientific Reports},
  year = {2025},
  volume = {15},
  number={1},
  pages = {16290},
  doi = {10.1038/s41598-025-01367-7}
}

@article{singh2025personalized,
  title={Personalized glucose prediction using in situ data only},
  author={Singh, Rohan and Toumi, Marouane and Salath{\'e}, Marcel},
  journal={Frontiers in Nutrition},
  volume={12},
  pages={1539118},
  year={2025},
  publisher={Frontiers Media SA},
  doi = {10.3389/fnut.2025.1539118}
}

@article{tominaga2025transformer,
  title={Prediction of Postprandial Blood Glucose Variability Using Machine Learning in Frequent Insulin Injection Therapy with a Simplified Carbohydrate Counting Model},
  author={Tominaga, Hiroyuki and Hamaguchi, Masahide and Hamaguchi, Youji and Yashiki, Ren and Yamaguchi, Aki and Arai, Tadaharu and Yamazaki, Masahiro and Kitagawa, Noriyuki and Hashimoto, Yoshitaka and Okada, Hiroshi and others},
  journal={Nutrients},
  volume={17},
  number={24},
  pages={3832},
  year={2025},
  doi = {10.3390/nu17243832}
}

@article{brugger2025ppg,
  title={Predicting postprandial glucose excursions to personalize dietary interventions for type-2 diabetes management},
  author={Br{\"u}gger, Victoria and Kowatsch, Tobias and Jovanova, Mia},
  journal={Scientific Reports},
  volume={15},
  number={1},
  pages={25920},
  year={2025},
  publisher={Nature Publishing Group UK London},
  doi={10.1038/s41598-025-08003-4}
}

@article{wang2025dynamic,
  author={Wang, Shihan and Song, Shuoning and Gao, Junxiang and Wu, Weiming and Fu, Yong and Yuan, Tao and Zhao, Weigang},
  title = {Dynamic prediction of postprandial glycemic response and personalized dietary interventions based on machine learning},
  journal = {The Journal of Nutrition},
  year = {2025},
  volume = {155},
  number = {12},
  pages = {4193-4208},
  doi = {10.1016/j.tjnut.2025.09.023}
}

@article{Chen2022GV,
  title={Long-term glycemic variability and risk of adverse health outcomes in patients with diabetes: A systematic review and meta-analysis of cohort studies},
  author={Chen, Junxiang and Yi, Qian and Wang, Yuxiang and Wang, Jingyi and Yu, Hancheng and Zhang, Jijuan and Hu, Mengyan and Xu, Jiajing and Wu, Zixuan and Hou, Leying and others},
  journal={Diabetes Research and Clinical Practice},
  volume={192},
  pages={110085},
  year={2022},
  publisher={Elsevier},
  doi={10.1016/j.diabres.2022.109123}
}

@article{Monnier2006OxStress,
  title={Activation of oxidative stress by acute glucose fluctuations compared with sustained chronic hyperglycemia in patients with type 2 diabetes},
  author={Monnier, Louis and Mas, Emilie and Ginet, Christine and Michel, Fran{\c{c}}oise and Villon, Laetitia and Cristol, Jean-Paul and Colette, Claude},
  journal={Jama},
  volume={295},
  number={14},
  pages={1681--1687},
  year={2006},
  publisher={American Medical Association},
  doi={10.1001/jama.295.14.1681}
}

@article{lin2024decision,
  title={Decision-oriented dialogue for human-AI collaboration},
  author={Lin, Jessy and Tomlin, Nicholas and Andreas, Jacob and Eisner, Jason},
  journal={Transactions of the Association for Computational Linguistics},
  volume={12},
  pages={892--911},
  year={2024},
  publisher={MIT Press 255 Main Street, 9th Floor, Cambridge, Massachusetts 02142, USA~…},
  doi={https://doi.org/10.1162/tacl_a_00679}
}

@inproceedings{lee2021mind,
  author    = {Changhun Lee and Soohyeok Kim and Sehwa Jeong and Chiehyeon Lim and Jayun Kim and Yeji Kim and Minyoung Jung},
  title     = {MIND dataset for diet planning and dietary healthcare with machine learning: Dataset creation using combinatorial optimization and controllable generation with domain experts},
  booktitle = {Proceedings of the Neural Information Processing Systems Track on Datasets and Benchmarks},
  year      = {2021},
  note      = {NeurIPS Datasets and Benchmarks 2021}
}

@inproceedings{lee2021diet,
  author    = {Changhun Lee and Soohyeok Kim and Chiehyeon Lim and Jayun Kim and Yeji Kim and Minyoung Jung},
  title     = {Diet Planning with Machine Learning: Teacher-forced REINFORCE for Composition Compliance with Nutrition Enhancement},
  booktitle = {Proceedings of the 27th ACM SIGKDD Conference on Knowledge Discovery and Data Mining},
  pages     = {3150--3160},
  year      = {2021},
  doi       = {10.1145/3447548.3467201}
}

@article{hettiarachchi2022integrating,
  author  = {Chirath Hettiarachchi and Elena Daskalaki and Jane Desborough and Christopher J. Nolan and David O'Neal and Hanna Suominen},
  title   = {Integrating Multiple Inputs Into an Artificial Pancreas System: Narrative Literature Review},
  journal = {JMIR Diabetes},
  volume  = {7},
  number  = {1},
  pages   = {e28861},
  year    = {2022},
  doi     = {10.2196/28861},
  pmid    = {35200143}
}

@article{tejedor2020reinforcement,
  author  = {Miguel Tejedor and Ashenafi Zebene Woldaregay and Fred Godtliebsen},
  title   = {Reinforcement learning application in diabetes blood glucose control: A systematic review},
  journal = {Artificial Intelligence in Medicine},
  volume  = {104},
  pages   = {101836},
  year    = {2020},
  doi     = {10.1016/j.artmed.2020.101836},
  pmid    = {32499004}
}

@inproceedings{Khamesian_2025,
  title        = {NutriGen: Personalized Meal Plan Generator Leveraging Large Language Models to Enhance Dietary and Nutritional Adherence},
  author       = {Khamesian, Saman and Arefeen, Asiful and Carpenter, Stephanie M. and Ghasemzadeh, Hassan},
  booktitle    = {2025 47th Annual International Conference of the IEEE Engineering in Medicine and Biology Society (EMBC)},
  year         = {2025},
  month        = jul,
  address      = {Copenhagen, Denmark},
  pages        = {1--7},
  publisher    = {IEEE},
  organization = {IEEE},
  doi          = {10.1109/EMBC58623.2025.11253879}
}

@article{zhao2023chinese,
  title={Chinese diabetes datasets for data-driven machine learning},
  author={Zhao, Qinpei and Zhu, Jinhao and Shen, Xuan and Lin, Chuwen and Zhang, Yinjia and Liang, Yuxiang and Cao, Baige and Li, Jiangfeng and Liu, Xiang and Rao, Weixiong and others},
  journal={Scientific Data},
  volume={10},
  number={1},
  pages={35},
  year={2023},
  publisher={Nature Publishing Group UK London}
}

@article{bent2021bigidea,
  title={Engineering digital biomarkers of interstitial glucose from noninvasive smartwatches},
  author={Bent, Brinnae and Cho, P and Henriquez, M and Wittmann, A and Thacker, C and Feinglos, M and Crowley, M and Dunn, Jessilyn},
  journal={NPJ Digital Medicine},
  volume={4},
  number={1},
  pages={89},
  year={2021},
  publisher={Nature Publishing Group},
  doi={10.1038/s41746-021-00465-w}
}

@misc{yang2025qwen3technicalreport,
      title={Qwen3 Technical Report}, 
      author={An Yang and Anfeng Li and Baosong Yang and Beichen Zhang and Binyuan Hui and Others},
      year={2025},
      eprint={2505.09388},
      archivePrefix={arXiv},
      primaryClass={cs.CL},
      url={https://arxiv.org/abs/2505.09388}, 
}

@article{hossain2025cgmacros,
  title   = {CGMacros: a pilot scientific dataset for personalized nutrition and diet monitoring},
  author  = {Das, Anurag and Kerr, David and Glantz, Namino and Bevier, Wendy and Santiago, Rony and Gutierrez-Osuna, Ricardo and Mortazavi, Bobak J.},
  journal = {Scientific Data},
  volume  = {12},
  number  = {1},
  pages   = {1557},
  year    = {2025},
  doi     = {10.1038/s41597-025-05851-7},
  publisher = {Nature Publishing Group}
}

@inproceedings{sergazinov2023gluformer,
  title={Gluformer: Transformer-Based Personalized Glucose Forecasting with Uncertainty Quantification},
  author={Sergazinov, Renat and Armandpour, Mohammadreza and Gaynanova, Irina},
  booktitle={2023 IEEE International Conference on Acoustics, Speech and Signal Processing (ICASSP)},
  pages={1--5},
  year={2023},
  organization={IEEE},
  doi={10.1109/ICASSP49357.2023.10096419}}

@article{farahmand2024gluconet,
  title={Hybrid Attention Model Using Feature Decomposition and Knowledge Distillation for Glucose Forecasting},
  author={Farahmand, Ebrahim and Soumma, Shovito Barua and Chatrudi, Nooshin Taheri and Ghasemzadeh, Hassan},
  journal={arXiv preprint arXiv:2411.10703},
  year={2024}
}

@article{li2020glunet,
  title={GluNet: A Deep Learning Framework for Accurate Glucose Forecasting},
  author={Li, Kezhi and Daniels, Jenna and Liu, Chang and Herrero, Pau and Georgiou, Pantelis},
  journal={IEEE Journal of Biomedical and Health Informatics},
  volume={24},
  number={2},
  pages={414--423},
  year={2020},
  publisher={IEEE}
}

@article{hochreiter1997long,
  title={Long short-term memory},
  author={Hochreiter, Sepp and Schmidhuber, J{\"u}rgen},
  journal={Neural Computation},
  volume={9},
  number={8},
  pages={1735--1780},
  year={1997},
  publisher={MIT Press}
}

@article{Dragomiretskiy2014VMD,
  title={Variational Mode Decomposition},
  author={Dragomiretskiy, Konstantin and Zosso, Dominique},
  journal={IEEE Transactions on Signal Processing},
  volume={62},
  number={3},
  pages={531--544},
  year={2014},
  publisher={IEEE},
  doi={10.1109/TSP.2013.2288675}
}

@article{Wang2020VMDLSTM,
  title={Blood Glucose Prediction With {VMD} and {LSTM} Optimized by Improved Particle Swarm Optimization},
  author={Wang, Wenbo and Tong, Meng and Yu, Min},
  journal={IEEE Access},
  volume={8},
  pages={217908--217916},
  year={2020},
  publisher={IEEE},
  doi={10.1109/ACCESS.2020.3041355}
}

@article{Hovorka2004MPC,
  title={Nonlinear model predictive control of glucose concentration in subjects with type 1 diabetes},
  author={Hovorka, Roman and Canonico, V and Chassin, LJ and Haueter, U and Massi-Benedetti, M and Orsini Federici, M and Pieber, TR and Schaller, HC and Schaupp, L and Vering, T and Wilinska, ME},
  journal={Physiological Measurement},
  volume={25},
  number={4},
  pages={905--920},
  year={2004},
  publisher={IOP Publishing},
  doi={10.1088/0967-3334/25/4/010}
}

@inproceedings{yang2024large,
  title={Large Language Models as Optimizers},
  author={Yang, Chengrun and Wang, Xuezhi and Lu, Yifeng and Liu, Hanxiao and Le, Quoc V and Zhou, Denny and Chen, Xinyun},
  booktitle={The Twelfth International Conference on Learning Representations},
  year={2024},
  url={https://openreview.net/forum?id=Bb4VGOWELI}
}

@inproceedings{yao2023react,
  title={{ReAct}: Synergizing Reasoning and Acting in Language Models},
  author={Yao, Shunyu and Zhao, Jeffrey and Yu, Dian and Du, Nan and Shafran, Izhak and Narasimhan, Karthik and Cao, Yuan},
  booktitle={The Eleventh International Conference on Learning Representations},
  year={2023},
  url={https://openreview.net/forum?id=WE_vluYUL-X}
}

@article{wu2025,
  author = {Wu, Yue and Ehlert, Ben and Metwally, Ahmed A. and Perelman, Dalia and Park, Heyjun and others},
  title = {Individual variations in glycemic responses to carbohydrates and underlying metabolic physiology},
  journal = {Nature Medicine},
  year = {2025},
  volume = {31},
  pages = {2232--2243},
  doi = {10.1038/s41591-025-03719-2}
}

@article{Metwally2025Prediction,
  title = {Prediction of metabolic subphenotypes of type 2 diabetes via continuous glucose monitoring and machine learning},
  author = {Metwally, Ahmed A. and Perelman, Dalia and Park, Heyjun and Wu, Yue and Jha, Alokkumar and Sharp, Seth and Celli, Alessandra and Ayhan, Ekrem and Abbasi, Fahim and Gloyn, Anna L. and McLaughlin, Tracey and Snyder, Michael P.},
  journal = {Nature Biomedical Engineering},
  volume = {9},
  number = {8},
  pages = {1222--1239},
  year = {2025},
  publisher = {Nature Publishing Group},
  doi = {10.1038/s41551-024-01311-6},
  url = {https://www.nature.com/articles/s41551-024-01311-6}
}

@article{Li2024RolePlay,
  title={ChatDiet: Empowering personalized nutrition-oriented food recommender chatbots through an LLM-augmented framework},
  author = {Yang, Zhongqi and Khatibi, Elahe and Nagesh, Nitish and Abbasian, Mahyar and Azimi, Iman and Jain, Ramesh and Rahmani, Amir M.},
  journal={Smart Health},
  volume={32},
  pages={100465},
  year={2024},
  publisher={Elsevier}
}

@inproceedings{du-etal-2025-llmagentSP,
    title = "{LLM}s Can Simulate Standardized Patients via Agent Coevolution",
    author = "Du, Zhuoyun  and
      LujieZheng, LujieZheng  and
      Hu, Renjun  and
      Xu, Yuyang  and
      Li, Xiawei  and
      Sun, Ying  and
      Chen, Wei  and
      Wu, Jian  and
      Cai, Haolei  and
      Ying, Haochao",
    editor = "Che, Wanxiang  and
      Nabende, Joyce  and
      Shutova, Ekaterina  and
      Pilehvar, Mohammad Taher",
    booktitle = "Proceedings of the 63rd Annual Meeting of the Association for Computational Linguistics (Volume 1: Long Papers)",
    month = jul,
    year = "2025",
    address = "Vienna, Austria",
    publisher = "Association for Computational Linguistics",
    url = "https://aclanthology.org/2025.acl-long.846/",
    doi = "10.18653/v1/2025.acl-long.846",
    pages = "17278--17306",
    ISBN = "979-8-89176-251-0"
}

@inproceedings{zhu-etal-2025-plangptllmagent,
  title={PlanGPT: Enhancing urban planning with a tailored agent framework},
  author={Zhu, He and Chen, Guanhua and Zhang, Wenjia},
  booktitle={Proceedings of the 63rd Annual Meeting of the Association for Computational Linguistics},
  pages={764--783},
  month={07},
  volume={6},
  year={2025},
  doi={10.18653/v1/2025.acl-industry.54}
}

@inproceedings{he-etal-2025-pasallmagent,
    title = "{P}a{S}a: An {LLM} Agent for Comprehensive Academic Paper Search",
    author = "He, Yichen  and
      Huang, Guanhua  and
      Feng, Peiyuan  and
      Lin, Yuan  and
      Zhang, Yuchen  and
      Li, Hang  and
      E, Weinan",
    editor = "Che, Wanxiang  and
      Nabende, Joyce  and
      Shutova, Ekaterina  and
      Pilehvar, Mohammad Taher",
    booktitle = "Proceedings of the 63rd Annual Meeting of the Association for Computational Linguistics (Volume 1: Long Papers)",
    month = jul,
    year = "2025",
    address = "Vienna, Austria",
    publisher = "Association for Computational Linguistics",
    url = "https://aclanthology.org/2025.acl-long.572/",
    doi = "10.18653/v1/2025.acl-long.572",
    pages = "11663--11679",
    ISBN = "979-8-89176-251-0"
}

@article{Marchal2025Applications,
  title = {Applications of digital twins in medicine},
  author = {Marchal, Iris},
  journal = {Nature Biotechnology},
  volume = {43},
  number = {10},
  pages = {1606--1612},
  year = {2025},
  publisher = {Nature Publishing Group},
  doi = {10.1038/s41587-025-02847-x},
  url = {https://www.nature.com/articles/s41587-025-02847-x}
}

\appendix
In this appendix, we provide the supplementary information accompanying the main paper, including additional data, explanations, and details.

\section{Experimental Detail}
\subsection{PAGP Training Parameters}
\label{app:pagp_hparams}

This appendix summarizes the training parameters used for the \textbf{Physiology-Aware Glucose Predictor (PAGP)}. PAGP predicts the low-frequency component of the postprandial glucose trajectory using an LSTM and the high-frequency component using a Transformer. Dietary and medication signals are first processed by the learnable \textbf{Temporal Physiological Absorption Decay Module (TPADM)}, while free-form meal descriptions are encoded by a BERT-style nutrition semantic encoder and then fused with structured inputs.

We adopt a subject-specific training strategy: for each subject, we first pretrain PAGP on the cohort from the same population group (e.g., healthy$\rightarrow$healthy, T1DM$\rightarrow$T1DM, T2DM$\rightarrow$T2DM), and then fine-tune the pretrained model on the target subject's training split. We select the best checkpoint by validation RMSE and apply early stopping. We keep the data split and random seed consistent across ablations.

We use a sliding input window of $T_{\mathrm{in}}=120$ minutes to predict the next $T_{\mathrm{out}}=120$ minutes of glucose, with a sampling interval determined by the corresponding dataset.  The low-frequency mode is modeled by the LSTM branch, and the high-frequency mode is modeled by the Transformer branch.
The LSTM branch uses a 2-layer unidirectional LSTM with hidden size 128 and dropout rate 0.2. The LSTM consumes the full fused structured input sequence over the input window (including the recent glucose history and other structured covariates), and outputs a low-frequency glucose prediction for the future horizon.
The Transformer branch is an encoder-style Transformer with 4 layers, model dimension $d_{\mathrm{model}}=128$, 4 attention heads, and feed-forward dimension 512. We use dropout 0.1 throughout the Transformer and employ a pre-layer-normalization (pre-LN) layout for stability. The Transformer input at each time step is formed by concatenating the TPADM outputs (diet and medication absorption/effect representations) with other structured features (e.g., exercise-related signals), followed by a linear projection to $d_{\mathrm{model}}$. The Transformer outputs a high-frequency glucose prediction over the future horizon using the same style of MLP regression head (128$\rightarrow$64$\rightarrow$1) with GELU and dropout 0.1.

TPADM outputs a time-resolved physiological effect vector with dimensionality $d_{\mathrm{tp}}=64$. We instantiate separate TPADM streams for dietary intake and medication action, and concatenate their outputs before fusion. All decay/absorption parameters are learnable and constrained to be non-negative using a softplus re-parameterization. TPADM is trained end-to-end jointly with the LSTM/Transformer predictors. Meal descriptions are encoded using a Chinese BERT-base style encoder with hidden size 768. We truncate the input to a maximum of 128 tokens and use the [CLS] representation as the pooled embedding. The pooled embedding is then mapped to the predictor feature space by an MLP projection (768$\rightarrow$128) with GELU activation and dropout 0.1, and is fused with structured inputs for both branches. 

We optimize PAGP with AdamW and use gradient clipping with $\ell_2$ norm threshold 1.0. We train with mixed precision (fp16) enabled. For the non-BERT modules (LSTM, Transformer, TPADM, fusion and heads), we use a learning rate of $1\times10^{-3}$ during pretraining and $5\times10^{-4}$ during subject-specific fine-tuning. For the BERT encoder, we use a smaller learning rate of $2\times10^{-5}$ during pretraining and $1\times10^{-5}$ during fine-tuning. Weight decay is set to 0.01 for all trainable parameters. We apply a cosine learning rate schedule with linear warmup. In pretraining, the warmup ratio is 5\% of total steps; in fine-tuning, the warmup ratio is 10\%. We train for up to 50 epochs in pretraining and up to 30 epochs in fine-tuning, and adopt early stopping based on validation RMSE with patience 8 (pretraining) and patience 6 (fine-tuning).

All models are implemented in PyTorch. We normalize structured inputs by z-score using training statistics only. Glucose values are trained in mg/dL; we report RMSE on the original scale. 

\subsection{PD-2SMO LLM Inference Defaults}
\label{app:pd2smo_llm_defaults}

PD-2SMO uses the following default LLM API and decoding parameters for all agent calls. We invoke an OpenAI-compatible Chat Completions API with a fixed instruction-tuned chat model (\texttt{model} is held constant within each experiment).
We set \texttt{temperature}$=0.2$, \texttt{top\_p}$=0.9$, \texttt{max\_tokens}$=2048$, \texttt{frequency\_penalty}$=0$, and \texttt{presence\_penalty}$=0$.
We use exactly one LLM call to propose a structured edit for the current iteration ($1$ call/iteration).
The LLM is instructed to produce a strict JSON object. If the response is not valid JSON, we perform one format-repair call using the same model with \texttt{temperature}$=0$ and \texttt{max\_tokens}$=2048$ to enforce schema compliance.

\subsection{PAGP Evaluation Metric}
\label{PAGPMetric}
The Root Mean Square Error (RMSE) is defined by the following equation:
\[
\text{RMSE} = \sqrt{\frac{1}{N} \sum_{i=1}^{N} \left(\hat{G}_i - G_i \right)^2 }.
\]

The the coefficient of determination ($R^2$) is defined by the following equation::
\[
R^2 = 1 - \frac{\sum_{i} (\hat{G}_i - G_i)^2}{\sum_{i} (G_i - \bar{G})^2}.
\]

\subsection{PD-2SMO Target}
Let $x_0$ denote the original meal and $x$ a candidate optimized meal. The relative improvement is defined as:
\begin{equation}
\mathrm{Imp}_{iAUC} =
\frac{iAUC^{x_0}_{2h} - iAUC^{x}_{2h}}{iAUC^{x_0}_{2h}}.
\end{equation}
The optimization target is $\mathrm{Imp}_{iAUC} \ge 10\%$ under practical dietary constraints.

\subsection{Dataset Preparation Details}

This appendix describes how we prepare the four datasets used in our experiments, including CGMacros, BIG IDEAS, Shanghai T1DM, and Shanghai T2DM. Although the raw modalities and logging completeness vary across datasets, we convert them into a unified subject-centric format with aligned glucose trajectories and structured event features for both prediction and intervention.

\subsubsection{Common Preprocessing Pipeline}
\label{app:common_pipeline}
For all datasets, we apply a consistent preprocessing pipeline.
(i) \textbf{Time alignment:} All timestamps are converted into a unified local timezone and aligned to the CGM sampling grid.
(ii) \textbf{CGM cleaning:} We remove implausible CGM values and discard segments with excessive missingness. Short gaps are imputed via linear interpolation, while long gaps result in segment truncation.
(iii) \textbf{Window construction:} We construct samples using a rolling-window strategy. Each sample contains a history context (CGM + events) and a future prediction horizon. Windows overlapping with severe missing CGM intervals are excluded.
(iv) \textbf{Data split:} We adopt subject-level splits whenever possible. Otherwise, we use within-subject chronological splits to avoid future leakage.

\subsubsection{CGMacros}
\label{app:cgmacros}
CGMacros provides meal records primarily in the form of food images. To integrate CGMacros into our text-conditioned prediction-intervention pipeline, we first convert each meal image into a textual recipe-style English description using an LLM-based captioning procedure.
Specifically, for each meal image, we prompt a vision-language model to produce (1) a concise dish description, (2) a structured ingredient list, and (3) an approximate portion estimate when possible. We then normalize the generated outputs into a standardized text recipe format and align each meal event to the closest CGM time step. If multiple images correspond to the same eating episode, we merge them into a single meal event by concatenating the generated ingredient lists and summing portion estimates when available. For quality control, we filter out captions that are empty, non-food, or clearly inconsistent with typical meal content, and we keep missing fields masked rather than forcing zero values. The resulting text recipes are used as the meal input to our model, enabling consistent handling of meal semantics across datasets.

\subsubsection{BIG IDEAS}
\label{app:bigideas}
BIG IDEAS includes multi-day CGM signals together with lifestyle records. We convert the raw logs into synchronized streams of CGM, meal events, and activity/medication events when available. We normalize all event times to the smallest available grid. We further enforce minimal quality constraints (e.g., minimum CGM coverage per day and minimum number of logged meals per subject) to reduce noise introduced by under-logging. The dietary logs in BIG IDEAS dataset are written in English.

\subsubsection{Shanghai T1DM and Shanghai T2DM}
\label{app:shanghai}
The Shanghai datasets contain CGM trajectories from clinical cohorts with type-1 diabetes (SH T1DM) and type-2 diabetes (SH T2DM), respectively. Compared with non-clinical datasets, Shanghai cohorts may contain richer medication records but limited lifestyle logging. Notably, \textbf{exercise/activity logs are not available} in the Shanghai datasets. Therefore, we remove the exercise modality from the input interface when training and evaluating on SH T1DM/SH T2DM.
Concretely, we keep the same CGM and meal/medication preprocessing steps as in other datasets, but we (i) omit the activity feature channels, and (ii) adjust the model input masks accordingly to ensure that missing exercise is not treated as observed zero activity. Medication entries (e.g., insulin and/or anti-diabetic drugs) are converted into time-stamped dose vectors, and meal records are processed as structured meal events when nutritional fields are present; otherwise, we retain timestamp-only meal markers as sparse indicators. The dietary logs in Shanghai T1DM and T2DM dataset are written in Chinese.

\subsubsection{Unified Feature Schema}
\label{app:unified_schema}
After preprocessing, all datasets are represented using a unified schema: each sample contains (a) a fixed-length CGM history segment, (b) synchronized event sequences (meal, medication, and activity when available) with modality-aware masks, and (c) a future CGM horizon for supervised learning. This unified representation enables training the Physiology-Aware Glucose Predictor (PAGP) and deploying the downstream meal optimization agent under consistent interfaces, while allowing modality drop-out (e.g., no-exercise) for datasets with incomplete logging.

\begin{table*}[htbp]
\small
\centering
\begin{tabular}{lccccccccc}
\toprule
\multirow{2}{*}{\textbf{Dataset}} & \multicolumn{3}{c}{\textbf{Original}} & \multicolumn{3}{c}{\textbf{ReAct}} & \multicolumn{3}{c}{\textbf{Ours}} \\
\cmidrule(lr){2-4} \cmidrule(lr){5-7} \cmidrule(lr){8-10}
 & \textbf{TIR} & \textbf{TAR} & \textbf{TBR} & \textbf{TIR} & \textbf{TAR} & \textbf{TBR} & \textbf{TIR} & \textbf{TAR} & \textbf{TBR} \\
\midrule
BIG IDEAS & 77.32 & 17.81 & 4.87 & 83.68 & 13.91 & 2.41 & \textbf{90.02} & \textbf{7.75} & \textbf{2.23} \\
CGMacros & 81.05 & 15.04 & 3.91 & 85.41 & 11.74 & 2.85 & \textbf{91.34} & \textbf{6.83} & \textbf{1.83} \\
Shanghai T1DM & 89.53 & 7.82 & 2.65 & 91.02 & 6.59 & 2.39 & \textbf{91.97} & \textbf{6.40} & \textbf{1.63} \\
Shanghai T2DM & 82.91 & 10.71 & 6.38 & 85.72 & 8.81 & 5.47 & \textbf{87.09} & \textbf{7.66} & \textbf{5.25} \\
\bottomrule
\end{tabular}%
\caption{Comparison of PPGR Regulation Metrics (TIR, TAR, TBR reported in \%) over a 120-minute horizon across four datasets.}
\label{tab:tir_metrics}
\end{table*}

\subsubsection{Data License Information}
We use the datasets strictly under their stated licenses and access policies, consistent with the intended use for research and education:
(i) \textbf{CGMacros} is released under \textit{CC BY-NC-SA 4.0} and is accessible provided users comply with the license terms.
(ii) \textbf{BIG IDEAs Data} is released under the \textit{Open Data Commons Attribution License v1.0} with an access policy that permits use as long as the license terms are followed.
(iii) \textbf{Shanghai T1DM and T2DM} is listed with \textit{CC BY 4.0}.
Accordingly, we provide proper attribution in the manuscript, use the data for non-commercial academic research, and do not impose additional restrictions beyond the original licenses.

\subsection{Rule-based feasibility filter formalization and implementation}
We apply a rule-based feasibility filter to prune infeasible candidates proposed by the meal optimization agent.
Let the original meal be $\mathcal{M}_0=\{(d_i,q_i^0)\}_{i=1}^{n}$ where $d_i$ is a dish (or ingredient item) and $q_i^0$ is its portion size.
A candidate meal is $\mathcal{M}=\{(d_i,q_i)\}_{i=1}^{n'}$ (possibly with substitutions in Stage~2).
Let $E(d,q)$ denote energy (kcal) contributed by item $d$ at portion $q$; then total energy is
\begin{equation}
E(\mathcal{M})=\sum_{(d,q)\in\mathcal{M}} E(d,q).
\end{equation}
We use indicator variables $s_i\in\{0,1\}$ to denote whether $d_i$ is substituted (only allowed in Stage~2), and $\Delta q_i=q_i-q_i^0$ for portion change.

\textbf{(1) Energy-fluctuation constraint.}
We bound the deviation from the baseline meal energy:
\begin{equation}
\label{eq:energy_bound}
\big|E(\mathcal{M})-E(\mathcal{M}_0)\big|\le \epsilon_E,
\end{equation}

In practice, we also apply a per-edit energy cap to prevent single-step spikes:
\begin{equation}
\label{eq:per_edit_energy}
\big|E(d_i,q_i)-E(d_i,q_i^0)\big|\le \epsilon_{E,i}\quad \forall i.
\end{equation}

\textbf{(2) Two-stage constraints (Stage~1 vs.\ Stage~2).}
Stage~1 only allows portion scaling without substitutions:
\begin{equation}
\label{eq:stage1}
\quad s_i=0,\ \forall i,\quad q_i \in \mathcal{Q}(q_i^0),
\end{equation}
where $\mathcal{Q}(q_i^0)$ is a discrete set of admissible portion levels (defined below).
Stage~2 allows substitution but keeps the edit budget bounded:
\begin{equation}
\label{eq:stage2}
\quad \sum_i s_i \le B_{\text{sub}},\quad \sum_i \mathbb{I}[\Delta q_i\neq 0] \le B_{\text{portion}}.
\end{equation}
To preserve the meal structure, a substituted dish $d_i$ must map to a compatible class $c(\cdot)$:
\begin{equation}
\label{eq:class_preserve}
c(d_i)=c(d_i') \ \ \text{for any substitution } d_i \rightarrow d_i',
\end{equation}
where $c(\cdot)$ can be defined by food groups (e.g., staple/protein/vegetable) or dish roles.

\textbf{(3) Dish-composition (ratio) constraints.}
Let $\mathcal{G}$ be a set of dish groups (e.g., staple, protein, vegetable), and $\mathcal{M}_g \subset \mathcal{M}$ be items in group $g$.
Define group energy (or weight) share
\begin{equation}
\label{eq:group_share}
r_g(\mathcal{M})=\frac{\sum_{(d,q)\in \mathcal{M}_g} E(d,q)}{\sum_{(d,q)\in \mathcal{M}} E(d,q)}.
\end{equation}
We enforce bounded shares and deviation from the original meal:
\begin{equation}
\label{eq:ratio_bounds}
\underline{r}_g \le r_g(\mathcal{M}) \le \overline{r}_g,\quad
\big|r_g(\mathcal{M})-r_g(\mathcal{M}_0)\big|\le \epsilon_{r,g}.
\end{equation}
Additionally, to avoid degenerate solutions dominated by a single component, we cap the maximum item contribution:
\begin{equation}
\label{eq:dominance}
\max_{(d,q)\in\mathcal{M}} \frac{E(d,q)}{E(\mathcal{M})} \le \tau.
\end{equation}

\textbf{(4) Granularity constraints.}
We restrict portion edits to discrete, realistic increments:
\begin{equation}
\label{eq:granularity}
q_i =\left\{ \max(0, q_i^0 + k\delta_q)\ \big|\ k\in\mathbb{Z},\ |k|\le K_i \right\},
\end{equation}
where $\delta_q$ is a fixed step size (e.g., $\delta_q=10$\,g or $0.25$ serving), and $K_i$ limits per-item adjustment range.
We further constrain the overall intervention magnitude by bounding the number of changed items and the total $\ell_1$ portion change:
\begin{equation}
\label{eq:l1_budget}
\sum_i \mathbb{I}[\Delta q_i\neq 0] \le B_{\text{chg}},\quad
\sum_i |\Delta q_i| \le B_{q}.
\end{equation}

\textbf{Implementation details.}
We implement the filter as a deterministic post-processor that takes a set of candidates produced by the agent and returns the feasible subset.
Given a candidate $\mathcal{M}$, we (i) compute $E(\mathcal{M})$ and all group shares $r_g(\mathcal{M})$; (ii) check stage-appropriate constraints (Eq.~\ref{eq:stage1} or Eq.~\ref{eq:stage2}); (iii) validate composition constraints (Eq.~\ref{eq:ratio_bounds}--\ref{eq:dominance}); and (iv) validate edit granularity and budgets (Eq.~\ref{eq:granularity}--\ref{eq:l1_budget}).
Candidates failing any rule are rejected.
If no candidate is feasible in Stage~1, the agent escalates to Stage~2. In practice, energy and ratio computations utilize the USDA FoodData Central and the Chinese Food Composition Table. During Stage 2, if the LLM hallucinates or proposes a food item that does not exist in database, the feasibility filter flags the energy and ratio constraints as unverifiable. Consequently, the candidate is automatically rejected, forcing the agent to regenerate a dietary proposal in the next iteration.

\section{Additional Result}
\subsection{PD-2SMO Constraint Analysis}
\label{const_anal}

Recipe rationality is assessed by an LLM-based evaluator, which judges whether the generated meal plans are nutritionally and logically reasonable. Removing the calorie constraint from the agent leads to a substantial degradation in recipe rationality across all datasets. Specifically, the proportion of reasonable recipe combinations drops from 91.34\% to 74.86\% on BIG IDEAS, 90.27\% to 73.97\% on CGMacros, 96.49\% to 79.25\% on Shanghai T1DM, and 97.81\% to 83.03\% on Shanghai T2DM. On average, this corresponds to an absolute decrease of approximately 15–18 percentage points, indicating that calorie-aware control plays a critical role in maintaining globally reasonable meal plans. The degradation becomes even more pronounced when both the calorie constraint and the ratio constraint are removed. In this setting, the rationality scores further decline to 63.90\%, 59.61\%, 70.37\%, and 71.58\% on BIG IDEAS, CGMacros, Shanghai T1DM, and Shanghai T2DM, respectively. Compared to the full agent configuration, this represents a total reduction of over 25–30 percentage points in some datasets, highlighting the importance of jointly enforcing caloric limits and macronutrient or dish-level ratio constraints.

These results demonstrate that the proposed agent does not merely rely on LLM generation capabilities, but critically depends on structured dietary constraints to guide the optimization process. Without explicit calorie and ratio restrictions, the agent tends to generate meal combinations that, while potentially diverse, are significantly less coherent and nutritionally reasonable according to LLM-based evaluation.

\subsection{Post-Optimization Time in Range Result}
\label{tir_result}
To provide a comprehensive clinical evaluation of long-term PPGR control, we additionally calculate the Time in Range (TIR), Time Above Range (TAR), and Time Below Range (TBR) for the predicted PPGR over the 120-minute horizon. The glucose range we selected for calculation is 80 to 180 mg/dL. As shown in Table \ref{tab:tir_metrics}, our proposed PD-2SMO consistently achieves superior PPGR control by significantly increasing the TIR while simultaneously reducing both TAR and TBR compared to the original meals and the ReAct baseline.

\section{Pseudo-code: PD-2SMO}

\begin{algorithm}[htbp]
\small
\caption{Prediction-Driven Two-Stage Meal Optimization Agent (PD-2SMO)}
\label{alg:gated_opro}
\begin{algorithmic}[1]
\REQUIRE
$x_0$, $u$; $S(u,x)$;
$R(x,\text{stage})$; $\epsilon=10\%$;
$\Pi$; $B$;
$\tau=10\%$.
\ENSURE
Optimized meal $X^\ast$.

\STATE $(A_0,G_0)\leftarrow S(u,x_0)$ \COMMENT{$A_0$: iAUC$_{2h}$, $G_0$: $\Delta G_{2h}$}
\STATE Initialize memory $\mathcal{M}$ with $(x_0,A_0,G_0)$
\STATE $c\leftarrow 1$ \COMMENT{simulation call counter}

\vspace{0.3em}
\STATE $\text{stage}\leftarrow\textsc{Dist}$

\WHILE{$c<B$}
    \STATE Construct prompt $P$ $(x_0,u,\mathcal{M},\text{stage})$
    \STATE Generate candidate set $C\sim\Pi(\cdot\mid P)$
    \FORALL{$x\in C$}
        \IF{$c\ge B$} \STATE \textbf{break} \ENDIF
        \IF{$R(x,\text{stage})=\textsc{FAIL}$}
            \STATE Update $\mathcal{M}$ with failure record
            \STATE \textbf{continue}
        \ENDIF
        \STATE $(A,G)\leftarrow S(u,x)$; $c\leftarrow c+1$
        \STATE Update $\mathcal{M}$ with $(x,A,G)$
        \IF{$(A_0-A)/A_0\ge\tau$}
            \STATE \textbf{return} $\textsc{SelectTiers}(\mathcal{M})$
        \ENDIF
    \ENDFOR
\ENDWHILE

\vspace{0.3em}
\STATE $\text{stage}\leftarrow\textsc{Sub}$

\WHILE{$c<B$}
    \STATE $x^\dagger\leftarrow\textsc{BestPlan}(\mathcal{M})$
    \STATE $t\leftarrow\textsc{IdentifyContributor}(x^\dagger)$
    \STATE $\mathcal{L}\leftarrow\textsc{RetrieveSubstitutes}(t)$
    \STATE Construct prompt $P$ $(x_0,u,\mathcal{M},\mathcal{L},\text{stage})$
    \STATE Generate candidate set $C\sim\Pi(\cdot\mid P)$
    \FORALL{$x\in C$}
        \IF{$c\ge B$} \STATE \textbf{break} \ENDIF
        \IF{$R(x,\text{stage})=\textsc{FAIL}$}
            \STATE Update $\mathcal{M}$ with failure record
            \STATE \textbf{continue}
        \ENDIF
        \STATE $(A,G)\leftarrow S(u,x)$; $c\leftarrow c+1$
        \STATE Update $\mathcal{M}$ with $(x,A,G)$
        \IF{$(A_0-A)/A_0\ge\tau$}
            \STATE \textbf{return} $\textsc{SelectTiers}(\mathcal{M})$
        \ENDIF
    \ENDFOR
\ENDWHILE

\STATE \textbf{return} $\textsc{SelectTiers}(\mathcal{M})$ \COMMENT{best-effort}
\end{algorithmic}
\end{algorithm}

\section{Pseudo-code: the ReAct Agent for Personalized Glucose Regulation}
\label{sec:baseline_react}

\begin{algorithm}[t]
\small
\caption{ReAct Agent Baseline for Meal-Level Glucose Regulation}
\label{alg:react_baseline}
\begin{algorithmic}[1]
\REQUIRE
$x_0$, $u$; $S(u,x)$;
$R(x,\text{stage})$; $\epsilon=10\%$;
$\Pi$; $B$;
$\tau=10\%$; action set $\mathcal{A}$.
\ENSURE
Optimized meal $X^\ast$.

\STATE $(A_0,G_0)\leftarrow S(u,x_0)$ \COMMENT{$A_0$: iAUC$_{2h}$, $G_0$: $\Delta G_{2h}$}
\STATE Initialize memory $\mathcal{M}$ with $(x_0,A_0,G_0)$
\STATE $c\leftarrow 1$ \COMMENT{simulation call counter}
\STATE $\text{stage}\leftarrow\textsc{ReAct}$
\STATE $x\leftarrow x_0$ \COMMENT{current feasible proposal}

\WHILE{$c<B$}
    \STATE Construct prompt $P(x_0,u,\mathcal{M},\text{stage})$
    \STATE $(\texttt{Thought},a)\sim \Pi(\cdot\mid P)$ \COMMENT{$a\in\mathcal{A}$ is a tool-like action}
    \IF{$a=\textsc{Stop}$}
        \STATE \textbf{break}
    \ENDIF
    \STATE $\tilde{x}\leftarrow \textsc{Execute}(a,x)$ \COMMENT{apply action with typed arguments}
    \IF{$R(\tilde{x},\text{stage})=\textsc{FAIL}$}
        \STATE Update $\mathcal{M}$ with failure record
        \STATE \textbf{continue}
    \ENDIF
    \STATE $(A,G)\leftarrow S(u,\tilde{x})$; $c\leftarrow c+1$
    \STATE Update $\mathcal{M}$ with $(\tilde{x},A,G)$
    \STATE $x\leftarrow \tilde{x}$
    \IF{$(A_0-A)/A_0\ge\tau$}
        \STATE \textbf{return} $\textsc{SelectTiers}(\mathcal{M})$
    \ENDIF
\ENDWHILE

\STATE \textbf{return} $\textsc{SelectTiers}(\mathcal{M})$ \COMMENT{best-effort under budget $B$}
\end{algorithmic}
\end{algorithm}

We implement a strong agentic baseline that directly applies the ReAct (Reason+Act) paradigm to personalized meal-level glucose regulation, using the same subject information and simulator interface as PD-2SMO.
The baseline differs from PD-2SMO in that it does \emph{not} enforce a two-stage structure (\textsc{Dist}$\rightarrow$\textsc{Sub}) nor candidate-set evaluation; instead, it performs a single ReAct loop that iteratively edits the meal via discrete tool-like actions and uses prediction feedback to decide the next action.

For each subject, the ReAct agent receives the same inputs $(x_0,u)$ as PD-2SMO:
(i) recent meal records (description, ingredients, quantities, timestamps),
(ii) medication and exercise logs,
(iii) a historical CGM window, and
(iv) constraints (e.g., caloric bounds, ingredient ratios, minimum staple intake).
We treat the personalized glucose predictor (PAGP) as the environment simulator $S(u,x)$:
given a meal proposal $x$, $S$ returns a predicted 120-minute glucose trajectory and its summary metrics $(A,G)$, where $A$ is iAUC$_{2h}$ and $G$ is $\Delta G_{2h}$.

At each iteration, the LLM samples a \emph{typed} action $a\in\mathcal{A}$ and applies it via $\textsc{Execute}(a,x)$ to obtain a new proposal $\tilde{x}$.
We use the same rule-based feasibility filter as PD-2SMO, i.e., $R(\tilde{x},\text{stage})$, to enforce hard constraints and reject invalid edits (logged into $\mathcal{M}$ as failures).
Concretely, $\mathcal{A}$ includes:
\begin{itemize}
    \item \textsc{AdjustPortion}(\textit{item}, \textit{scale}) to scale grams/servings;
    \item \textsc{Substitute}(\textit{item}, \textit{replacement}) for ingredient replacement;
    \item \textsc{RecomposeMeal}(\textit{components}) to rebalance meal components;
    \item \textsc{Stop} to terminate the loop.
\end{itemize}

We use the simulator-call budget $B$ with counter $c$ and maintain a unified memory $\mathcal{M}$ of evaluated proposals.
The baseline terminates when (i) it outputs \textsc{Stop}, (ii) the budget is exhausted ($c\ge B$), or (iii) it achieves the same regulation threshold as PD-2SMO:
\begin{equation}
\frac{A_0-A}{A_0}\ge \tau \quad (\tau=10\%),
\end{equation}
where $(A_0,G_0)$ is obtained by simulating the original meal $x_0$.

\end{document}